\documentclass{emulateapj}

\newcommand{\be}{\begin{enumerate}}
\newcommand{\ee}{\end{enumerate}}

\newcommand{\logoh}{$\log(\mbox{O}/\mbox{H})$}
\newcommand{\tlogoh}{$12+\log(\mbox{O}/\mbox{H})$}

\shorttitle{Metal-Rich Dwarf Galaxies}
\shortauthors{Peeples, Pogge, \& Stanek}

\begin{document}

\title{Outliers from the Mass--Metallicity Relation I: A Sample of Metal-Rich
Dwarf Galaxies from SDSS}

\author{Molly S.\ Peeples, Richard W.\ Pogge, \& K.\ Z.\ Stanek}
    \affil{Department of Astronomy, Ohio State University, 140 W.\ 18th
    Ave., Columbus,~OH~43210,\\ molly@astronomy.ohio-state.edu, pogge@astronomy.ohio-state.edu, kstanek@astronomy.ohio-state.edu}

\begin{abstract}
We have identified a sample of 41 low-mass high--oxygen abundance
outliers from the mass--metallicity relation of star-forming galaxies
measured by \citet{tremonti04}.  These galaxies, which have $8.6 <
12+\log(\mbox{O}/\mbox{H}) < 9.3$ over a range of $-14.4 > M_B > -19.1$
and $7.4 < \log (\mbox{M}_{\star}/\mbox{M}_{\odot}) < 10$, are
surprisingly non-pathological.  They have typical specific star
formation rates, are fairly isolated and, with few exceptions, have
no obvious companions.  Morphologically, they are similar to dwarf
spheroidal or dwarf elliptical galaxies.  We predict that their observed
high oxygen abundances are due to relatively low gas fractions, concluding
that these are transitional dwarf galaxies nearing the end of their star
formation activity.
\end{abstract}


\section{Introduction}\label{sec:intro}
There is a well-known positive correlation between galaxy luminosity and
metallicity \citep{lequeux79, garnett87}.  \citet{tremonti04} measured
this relation for 53400 star-forming galaxies from the Sloan Digital Sky
Survey (SDSS) Data Release 4 (DR4, \citealt{adelman06}), estimating the
gas phase oxygen abundance, \tlogoh, from \ion{H}{2} region emission
lines as the surrogate for ``metallicity.''  Using estimates of the
galaxy stellar masses derived from the techniques of
\citet{kauffmann03}, \citeauthor{tremonti04} showed that metallicity is
better correlated with galaxy mass than with luminosity, which suggests
that the observed relation arises from the fact that lower mass galaxies
have lower escape velocities than higher mass galaxies and so lose
metals more easily via, e.g., supernova winds \citep{larson74}.
However, \citet{dalcanton07} showed that even if a low-mass galaxy can
selectively remove its metals via winds, subsequent star formation can
essentially erase the effects on the gas-phase metallicity; both a low
star formation rate and a large gas fraction are needed to retain low
metal abundances.  Other proposed origins for the mass--metallicity
relation include lower star formation rates in less massive galaxies
giving rise to fewer massive stars and therefore less metal production
\citep{koppen07}.

\begin{figure*}
\plotone{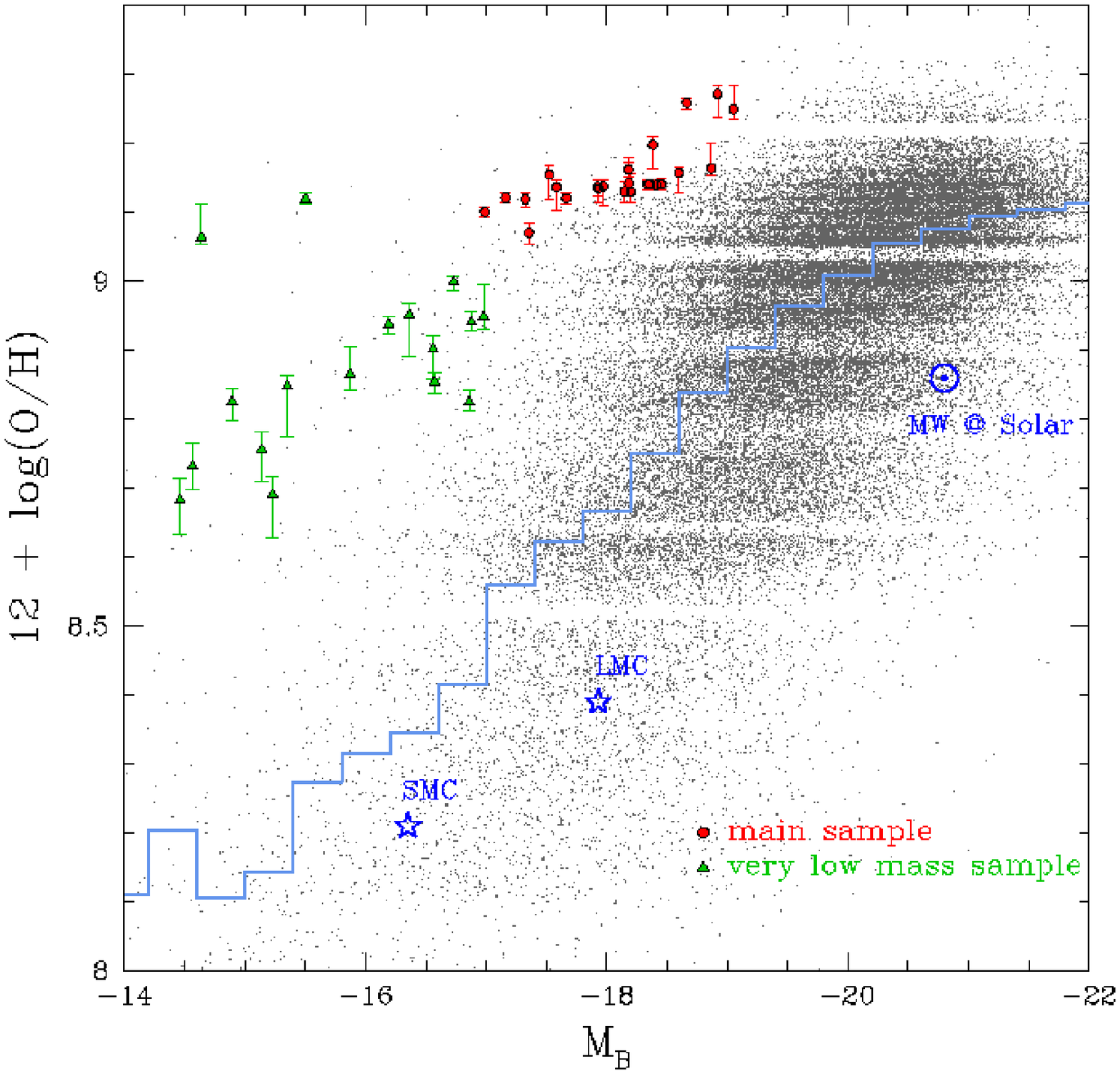}
\caption{\label{fig:ohmb} \tlogoh\ vs.\ $M_B$, with \logoh\ taken from
  \citet{tremonti04} and $M_B$ measured using the methods of
  \citet{assef08}.  The small grey points are star-forming galaxies with
  SDSS magnitude errors $<0.1$~mag; the red circles are the 24 galaxies
  in our main sample, and the green triangles are the very low mass
  sample.  The errorbars on \tlogoh\ show the central 68\% spread from
  \citet{tremonti04}.  The pale blue histogram denotes the median
  \tlogoh\ in bins of $M_B$ with width $\Delta M_B = 0.4$~mag; these
  medians are probably artificially high at low luminosities due to
  substantial contamination at high \tlogoh\ as discussed in
  \S\,\ref{sec:lowm}.  The Milky Way, SMC, and LMC are shown using $M_B$
  from \citet{karachentsev05}, and the \tlogoh\ measurements are as
  discussed in the \S\,\ref{sec:outliers}. }
\end{figure*}

Despite the locus of galaxies in the mass--metallicity plane having
intrinsically high scatter (e.g., a 1-$\sigma$ spread of $\pm 0.15$~dex
in \tlogoh\ at $9.5 < \log(\mbox{M}_{\star}/\mbox{M}_{\odot}) < 9.6$),
in the absence of good quality spectra it has become an increasingly
common practice to deduce or assign metallicities to galaxies based
simply on their luminosities.  In particular, the idea persists that
dwarf galaxies are always ``metal-poor,'' certainly a fair assumption in
the Local Group.  In this study however, we will show that there is a
significant population of low-mass galaxies with high gas-phase oxygen
abundances relative to the mass--metallicity relation defined by
\citet{tremonti04}; high-mass low-metallicity outliers from the
mass--metallicity relation will be the focus of an upcoming paper.  As
shown in Figure~\ref{fig:ohmb}, we find 24 galaxies with $-17 \gtrsim
M_B \gtrsim -19$~mag and $12 + \log(\mbox{O}/\mbox{H} \gtrsim 9$ and 17
high-metallicity outliers with $M_B > -17$~mag.  We explain how we
selected this sample in \S\,\ref{sec:outliers} and verified the
galaxies' high oxygen abundances in \S\,\ref{sec:metal}.  In
\S\,\ref{sec:disc} we discuss some of the possible explanations for the
existence of such a population of galaxies and explain why we predict
that these galaxies should have relatively low gas masses.
Specifically, in \S\,\ref{sec:trans} we discuss the evidence for our
galaxies being so-called ``transition'' dwarf galaxies. Finally, we
summarize our results and conclusions in \S\,\ref{sec:conc}.

\section{Finding Outliers from the Mass--Metallicity Relation}\label{sec:outliers}
We began with the sample of $\sim 110000$ star-forming galaxies with
measured gas-phase oxygen abundances and stellar masses from
\citet{tremonti04}.  High-metallicity outliers from the
mass--metallicity locus can be due to main three causes, and our cuts
were taken with these in mind. First, some galaxies can spuriously
appear to be underluminous for their mass.  Examples include highly
inclined galaxies subject to strong internal extinction and \ion{H}{2}
regions in larger galaxies that SDSS has mistakenly flagged as a galaxy.
Among relatively nearby galaxies (e.g., those in the Virgo cluster) some
have large peculiar velocities relative to the Hubble flow, leading to a
misestimation of their distance modulus and thus absolute magnitude.
Secondly, rogue outliers can have spuriously high estimated
metallicities; the galaxy, for example, may not be at high enough
redshift to have the strongly constraining [\ion{O}{2}]\,$\lambda
3727,9\,$\AA\ emission line in the SDSS bandpass.  Also, the strengths of
the [\ion{O}{2}] and [\ion{O}{3}] lines paradoxically {\em decrease}
with oxygen abundance, so at low signal-to-noise, the metallicity can be
overestimated due to underestimated line fluxes.  Finally, the object
can be an honest outlier; these are the objects we want in our final
sample.  We found that a large number of cuts in different parameter
spaces is useful for automatically rejecting many objects which would
otherwise have to be thrown out by visual inspection.  We ran two
searches on the full \citeauthor{tremonti04}\ sample for
high-metallicity outliers.  For the first, we were very conservative
with our cuts so as to be certain that the remaining galaxies are both
statistically significant and not spurious.  However, due to significant
contamination at low luminosities, this ``main'' sample has no members
with $\log\mbox{M}_{\star} < 9.15$.  We therefore did a separate, less
stringent search for the lowest luminosity outliers; to distinguish it
from the main sample, we refer to this sample of lower mass galaxies as
the ``very low mass'' sample.

\begin{figure*}
\plotone{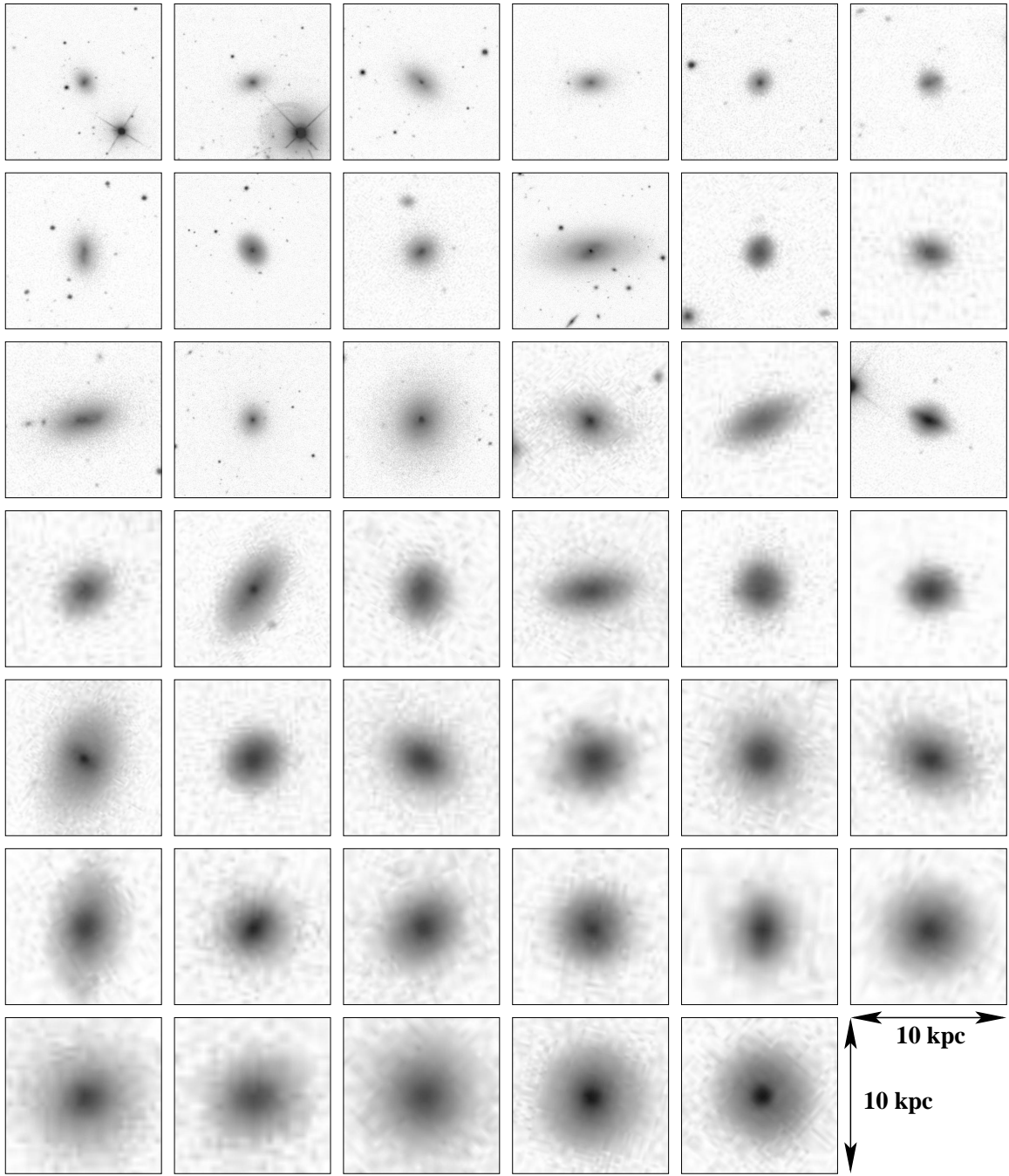}
\caption{\label{fig:images}SDSS images of high-metallicity dwarf
  galaxies scaled to $10\times10$\,kpc, with $M_B$ decreasing to the
  right and down from $-14.4$ to $-19.1$, so that the faintest galaxy is
  in the upper-left corner.  The stellar masses for this sample range
  from $\log\mbox{M}_{\star} \approx 7.4$ to 9.9.}
\end{figure*}

Figure~\ref{fig:ohmb} shows where our sample falls in the \tlogoh--$M_B$
plane; all galaxy images are shown in Figure~\ref{fig:images} and
summary information is presented in Table~\ref{tbl:sample}.  For
reference, in Figure~\ref{fig:ohmb} we plot the Small and Large
Magellanic Clouds and Milky Way using $M_B$ from \citet{karachentsev05}.
We recalculated \tlogoh\ for the SMC and LMC using the line ratios
reported by \citet{russell90} of six \ion{H}{2} regions in the SMC and
four \ion{H}{2} regions in the LMC and the relation $12 +
\log(\mbox{O}/\mbox{H}) = 9.37 + 2.03\times \mbox{N}2 +
1.26\times\mbox{N}2^2 + 0.32\times\mbox{N}2^3$, where $\mbox{N}2 \equiv
\log([\mbox{\ion{N}{2}}]\lambda6584/\mbox{H}\alpha)$
\citep{pettini04,kewley08}.  We then transformed these abundances onto
the \citet{tremonti04} scale using the formulae given by
\citet{kewley08}, resulting in average metallicities of 8.21 and 8.39
for the SMC and LMC, respectively.  The Milky Way is plotted also
plotted for reference at the Solar oxygen abundance of 8.86
\citep{delahaye06}.\footnote{We note that, unlike the other abundances
plotted in Figure~\ref{fig:ohmb}, the Solar abundance is neither a
nebular abundance nor the abundance within the central $\sim 5\,$kpc of
the Galaxy.}

\begin{deluxetable*}{rrclcrl}
\tabletypesize{\scriptsize}
\tablecaption{Full sample of low-luminosity mass--metallicity outliers
\label{tbl:sample} }
\tablecolumns{6}
\tablehead{
\colhead{RA} &
\colhead{dec} &
\colhead{\tlogoh} &
\colhead{$M_B$} &
\colhead{$\log \mbox{M}_{\star}$} &
\colhead{redshift} &
\colhead{Notes}}
\startdata
193.9056 & $-1.32986$ & 8.68 & $-14.46$ & 7.65 & 0.0029 & blue core, IC~225\\ 
180.4589 & $55.14507$ & 8.73 & $-14.56$ & 7.73 & 0.0035 & \\ 
190.2097 & $4.52583$ & 9.06 & $-14.64$ & 7.39 & 0.0025 & blue core, VCC~1855\\ 
179.0295 & $64.35073$ & 8.83 & $-14.89$ & 8.06 & 0.0047 & \\ 
228.0340 & $1.58571$ & 8.76 & $-15.14$ & 8.21 & 0.0065 & \\ 
126.6407 & $25.49979$ & 8.69 & $-15.23$ & 8.07 & 0.0072 & \\ 
227.2679 & $0.82197$ & 8.85 & $-15.35$ & 8.17 & 0.0055 & \\ 
193.6735 &  $2.10447$ & 9.12 & $-15.50$ & 8.40 & 0.0029 & bright core \\ 
126.6633 & $25.59821$ & 8.86 & $-15.87$ & 8.59 & 0.0078 & \\ 
208.3607 & $5.20778$ & 8.94 & $-16.19$ & 8.64 & 0.0027 & blue core \\ 
128.5841 & $50.45248$ & 8.95 & $-16.36$ & 8.69 & 0.0114 & \\ 
40.3408 & $0.05813$ & 8.90 & $-16.56$ & 8.70 & 0.0227 & \\ 
212.9768 & $53.93956$ & 8.85 & $-16.57$ & 8.81 & 0.0064 & blue core\\ 
190.5886 & $2.06662$ & 9.00 & $-16.73$ & 8.40 & 0.0044 & bright core\\ 
36.6179 & $1.16053$ & 8.82 & $-16.86$ & 7.92 & 0.0051 & blue core \\ 
117.1775 & $26.53979$ & 8.94 & $-16.88$ & 9.08 & 0.0155 & \\ \hline
182.7831 & $0.95682$ & 8.95 & $-16.98$ & 9.13 & 0.0206 & \\ 
133.8883 & $31.21168$ & 9.10 & $-16.99$ & 9.29 & 0.0068 & \\ 
25.7673 & $14.52434$ & 9.12 & $-17.16$ & 9.25 & 0.0284 & \\ 
225.5670 & $38.80631$ & 9.12 & $-17.32$ & 9.45 & 0.0147 & bright core \\ 
142.9797 & $39.27156$ & 9.07 & $-17.36$ & 9.35 & 0.0275 & \\ 
139.3797 & $33.47552$ & 9.15 & $-17.52$ & 9.46 & 0.0221 & \\ 
202.2088 & $-0.90846$ & 9.14 & $-17.58$ & 9.26 & 0.0217 & \\ 
39.0487 & $-7.73400$ & 9.12 & $-17.66$ & 9.16 & 0.0314 & \\ 
143.4695 & $41.07965$ & 9.14 & $-17.92$ & 9.54 & 0.0145 & bright core \\ 
187.1760 & $44.09813$ & 9.13 & $-17.93$ & 9.35 & 0.0239 & \\ 
182.8574 & $44.43604$ & 9.14 & $-17.97$ & 9.52 & 0.0231 &\\ 
211.3199 & $54.20347$ & 9.13 & $-18.15$ & 9.54 & 0.0417 & \\ 
258.4301 & $57.18840$ & 9.16 & $-18.18$ & 9.61 & 0.0290 & \\ 
208.9341 & $4.24367$ & 9.14 & $-18.19$ & 9.68 & 0.0295 & \\ 
227.6711 & $41.16220$ & 9.13 & $-18.20$ & 9.61 & 0.0316 & \\ 
352.8635 & $13.90885$ & 9.14 & $-18.34$ & 9.74 & 0.0324 & \\ 
206.4402 & $63.85402$ & 9.14 & $-18.36$ & 9.58 & 0.0313 & \\ 
153.2177 & $12.34451$ & 9.20 & $-18.39$ & 9.57 & 0.0315 & \\ 
177.7870 & $49.69415$ & 9.14 & $-18.41$ & 9.41 & 0.0482 & \\ 
144.1848 & $33.91996$ & 9.14 & $-18.45$ & 9.71 & 0.0426 & \\ 
162.6325 & $0.36061$ & 9.16 & $-18.60$ & 9.68 & 0.0384 & \\ 
195.6390 & $-3.33894$ & 9.26 & $-18.66$ & 9.55 & 0.0471 & \\ 
235.5660 & $51.73211$ & 9.16 & $-18.87$ & 9.83 & 0.0425 & \\ 
163.2191 & $43.42840$ & 9.27 & $-18.92$ & 9.90 & 0.0242 & \\ 
245.3673 & $40.20360$ & 9.25 & $-19.05$ & 9.91 & 0.0285 & blue core \\ 
\enddata
\tablecomments{Sample of metal-rich dwarf galaxies, sorted by
  $M_B$. First sixteen lines (above the horizontal line) are the very
  low mass ($\log [\mbox{M}_{\star}/\mbox{M}_{\odot}] < 9.1$) sample.
  RA and dec are in degrees, \tlogoh\ and stellar mass from
  \citet{tremonti04}, and $M_B$ is measured using the low-resolution
  templates of \citet{assef08} and corrected where necessary for
  peculiar velocities, as discussed in \S\,\ref{sec:outliers}; redshifts
  are spectroscopic redshifts and have not been corrected for peculiar
  velocities.}
\end{deluxetable*}

\subsection{Main Sample Selection}\label{sec:main}
\begin{figure}
\plotone{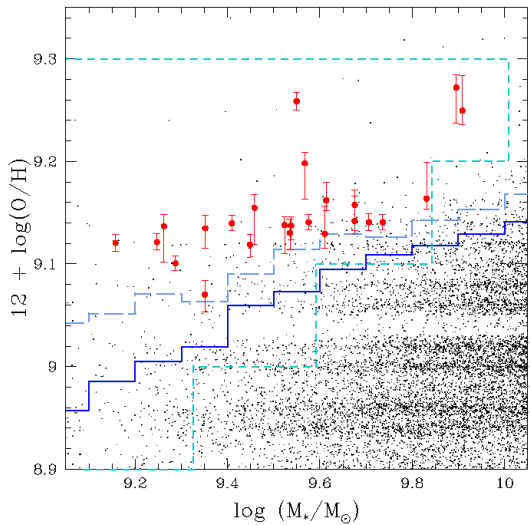}
\caption{\label{fig:cuts} Zoomed in portion of the \tlogoh\ vs.\
  $\log\mbox{M}_{\star}$ plane with 97.5\% (solid blue line and short
  dashed cyan line) and 99\% (long dashed light blue line) cuts shown;
  see \S\,\ref{sec:main} and Table~\ref{tbl:cuts}.  The black points
  have SDSS magnitude errors $<0.1$~mag and the red points are the 24
  galaxies in our main sample, which are similar in luminosity to the
  Large Magellanic Cloud (which has a \tlogoh\ of 8.39).  The errorbars
  on \tlogoh\ show the central 68\% spread from \citet{tremonti04}.}
\end{figure}

Table~\ref{tbl:cuts} summarizes our selection of the main sample of 24
metal-rich low-mass galaxies.  Spiral galaxies are known to have radial
metallicity gradients such that the nuclear abundances are higher than
averages over whole galaxies. Thus, if only a small fraction of the
galaxy is within the SDSS spectroscopic $3\arcsec$\ diameter fibers, the
measured metallicity can appear to be artificially high relative to
other galaxies.  Following \citet{tremonti04}, we make an initial cut by
requiring that the fraction of the galaxy covered by the fiber to be
greater than 10\%.  (In our final cut, we follow \citet{michel08} and
change this lower bound to 20\%, which only eliminates two of the
galaxies contained in the penultimate sample listed in
Table~\ref{tbl:cuts}.)  We wanted the sample to be statistically
significant, so outliers were then determined via a series of cuts in
the \tlogoh\ versus absolute $B$-band magnitude ($M_B$), $g$-band
($M_g$), and stellar mass (M$_{\star}$) planes, as demonstrated with
M$_{\star}$ in Figure~\ref{fig:cuts}.  For example, we divided the 52477
objects with SDSS magnitude errors $< 0.1$~mag in all bands into bins of
$M_B$ of width $\Delta M_B = 0.4$~mag; in each bin we kept the 2.5\%
with the highest \tlogoh.  We likewise took bins of \tlogoh\ of width
0.1~dex and kept 2.5\% of the objects with the faintest $M_B$.  Similar
cuts were made with $M_g$ (binsize $\Delta M_g = 0.4$~mag) and
$\log\mbox{M}_{\star}$ (binsize of $\Delta
\log\mbox{M}_{\star}=0.1$~dex).  $M_g$ was calculated from the SDSS
$g$-band magnitude and the spectroscopic redshift, $M_B$ was calculated
using the low-resolution spectral templates of \citealt{assef08}, and
the stellar masses from the \citeauthor{tremonti04}\ sample were
measured using a combination of SDSS colors and spectra as described by
\citet{kauffmann03}.  Only 58 objects survived this series of cuts.

At this point, we made a redshift cut: galaxies must have $z > 0.024$ in
order to have the [\ion{O}{2}]~$\lambda\lambda 3727,9$\AA\ emission line
pair in their SDSS spectrum, and as this is a highly constraining line
for the metallicity, we determined that it should be in the spectrum in
order to remove a possible source of systematics and so that we can
measure the metallicity with a diagnostic that uses this line.  Seven $z
< 0.024$ galaxies pass our subsequent \tlogoh\ error and visual
inspection cuts; after studying their spectra, as discussed below, we
chose to keep these galaxies in our main sample.  We also forced the
cited $\pm 1\,\sigma$ error in \tlogoh\ to be less than 0.05~dex;
metallicities with large uncertainties are obviously more likely to be
spurious than ones with small uncertainties.  Unsurprisingly, the $z >
0.024$ redshift cut (after all of the other cuts) also removed any
galaxies with fiber fractions $<0.1$ which would have survived to that
stage.  We find, however, that this large number of cuts in various
parameter spaces dramatically reduced the number of objects that had to
be removed ``by eye.''  The final visual-inspection cut removed only two
objects with nearby potential photometric contaminations.  Finally, we
re-examined the fiber fraction cut; following \citet{michel08} we
allowed the lower-limit fiber fraction to be 20\% (excluding two objects
from the sample).  The final exclusion was a barred galaxy with a
diameter of $\sim 10$\arcsec; the SDSS fiber size is $3\arcsec$, but
this galaxy is labeled as having a fiber fraction of $\sim 60$\%.  This
galaxy probably has spuriously low SDSS Petrosian magnitudes.  The 24
galaxies in this main sample have a range of fiber fractions between
25\% and 60\% and a redshift range of $0.007 < z < 0.048$.

\begin{deluxetable*}{ll}
\tablecaption{Cuts for Main Sample Selection
\label{tbl:cuts} }
\tablecolumns{2}
\tablehead{
\colhead{Cut} &
\colhead{Number Surviving}}
\startdata
Fiber fraction $> 0.1$				   & 107992\\	    
$k$-corrected SDSS magnitude errors $< 0.1$~mag    &  48327\\
97.5\% large (O/H) and small $M_B$ 		   &    227\\	    
97.5\% large (O/H) and small $M_g$ 		   &    202\\	    
99\% large (O/H) w.r.t.\ $M_B$ 			   &     87\\	    
99\% large (O/H) w.r.t.\ M$_{\star}$		   &	 66\\	    
97.5\% small M$_{\star}$ w.r.t.\ (O/H)		   &	 51\\	    
redshift $> 0.024$ for [\ion{O}{2}]			   &	 34\\	    
\tlogoh\ error $< 0.05$~dex			   &     22\\	    
Visual inspection                                  &	 20\\
Fiber fraction $> 0.2$ and photometry              &	 17\\ \hline
redshift $z < 0.024$ but surviving subsequent cuts &     $7$ 	    
\enddata
\tablecomments{See \S\,\ref{sec:main} and Figure~\ref{fig:cuts} for a
  more detailed explanation.
}
\end{deluxetable*}

\subsection{Very Low Mass Sample Selection}\label{sec:lowm}
It is obvious from Figure~\ref{fig:ohmb} that the selection of our main
sample artificially imposes a lower mass cutoff.  
This does not mean that there is not a statistically significant sample
of interesting galaxies with $M_B > -17$; it just means that there are
more contaminating objects with measured high metallicities in the
low-luminosity regime than at brighter magnitudes.  That is, the most
extreme outliers with $M_B < -17$ in one parameter space (e.g., the
\tlogoh\ vs.\ $M_B$ plane) are not also outliers in one of the others
(e.g., the \tlogoh\ vs.\ $\log\mbox{M}_{\star}$ plane). The {\em true}
very low luminosity mass--metallicity outliers therefore have lower
abundances than these spurious outliers.
  We therefore did a
separate search for high-metallicity galaxies at extremely low
luminosities and masses, as summarized in Table~\ref{tbl:lowmcuts}.  We
chose to not apply a fiber fraction cut for this sample because lower
luminosity galaxies are preferentially closer and therefore subtend a
larger angle on the sky, making it more difficult for a substantial
fraction of the galaxy to be within the 3\arcsec\ SDSS fiber diameter.
We therefore began the selection with a magnitude error cut, like the
one for the main sample.  We also removed objects with fiber fractions
$>0.2$ occupying the parameter space already excluded by the main
sample: any objects with $M_B$ brighter than $-17$~mag,
$\log\mbox{M}_{\star}>9.15$, and a high \logoh\ at the 95\% level with
respect to M$_{\star}$ were excluded.  The cuts in \tlogoh\ relative to
$M_B$, $M_g$ and M$_{\star}$ are all less stringent than for the main
sample, but because for this sample we are interested in the {\em very}
low mass objects, we took a strong (99\% level) cut in M$_{\star}$
relative to \tlogoh.  Six galaxies were excluded due to metallicity
re-estimation considerations (as discussed below).  Three of the
remaining galaxies are potentially members of the Virgo cluster; of
these, only two remain clear outliers from the $M_B$--metallicity locus
when their distance moduli are shifted to the Virgo value of
$30.74$~magnitudes \citep{ebeling98}.  (We do not attempt to correct the
stellar mass estimates due to the effects of peculiar velocity.)
Finally, the 95\% high (O/H) with respect to $M_g$ cut excludes the
galaxy IC~225 from the sample, which is pointed out by \citet{gu06} to
have a relatively high metallicity and a compact blue core.  As this
galaxy passes all of the other cuts and is clearly interesting, we added
it back to the sample, raising the total number of galaxies in the very
low mass sample to 17 and the full sample to 41.  The very low mass
sample has fiber fractions ranging from 0.04~to~0.45; more than half of
the 17 galaxies in this sample have fiber fractions below 0.1 and only 5
are above 0.2.  Also, none of these galaxies would have survived the
main sample redshift cut, as the very low mass galaxies occupy the
redshift range $0.00249 < z < 0.0227$, with the most nearby objects
being only $\sim 12$~Mpc away.

\begin{deluxetable*}{ll}
\tablecaption{Cuts for Very Low Mass Sample Selection
\label{tbl:lowmcuts} }
\tablecolumns{2}
\tablehead{
\colhead{Cut} &
\colhead{Number Surviving}}
\startdata
$k$-corrected SDSS magnitude errors $< 0.1$~mag    &  52744\\	    
84\% high (O/H) w.r.t. $M_B$                       &  8405\\
$M_B > -17$~mag                                    &  217\\
$\log\mbox{M}_{\star}<9.15$     &  201\\
84\% high (O/H) w.r.t. M$_{\star}$                 &  192\\
\tlogoh\ error $< 0.05$~dex		           &  139\\
99\% low $\log\mbox{M}_{\star}$ w.r.t. (O/H)       &  105\\
{\em exclude} 95\% high (O/H) w.r.t. M$_{\star}$, fiber fraction $>0.1$   &   78\\
95\% high (O/H) w.r.t. $M_g$                       &   32\\
Visual inspection                                  &   23\\
Metallicity comparison                             &   17\\
Absolute magnitude correction                      &   16
\enddata
\tablecomments{See \S\,\ref{sec:lowm} and Figure~\ref{fig:cuts} for
a more detailed explanation.  The 95\% high (O/H) w.r.t. $M_g$ cut excludes
IC~225, which passes subsequent cuts and is nonetheless included in
plots and discussions.  Note lack of fiber-fraction cut.  }
\end{deluxetable*}

\subsection{Measuring High Oxygen Abundances}\label{sec:metal}
Are these 41 galaxies true outliers from the mass--metallicity relation,
or are they just the tail of the scatter of the \citet{tremonti04}
measurements?  As a first check, the galaxies fall where expected on the
standard \citeauthor*{baldwin81} (BPT) diagrams for
$\log(\mbox{[\ion{O}{3}]}\,\lambda 5007/\mbox{H}\beta)$ vs.\
$\log(\mbox{[\ion{N}{2}]}\,\lambda 6548/\mbox{H}\alpha)$ and
$\log(\mbox{([\ion{S}{2}]}\,\lambda\lambda 6717+31)/\mbox{H}\alpha)$
\citep{baldwin81,kewley06b}. We note that there are several
complications with measuring high metallicities using visible wavelength
spectra (see \citealt{bresolin06} for a thorough review).  The main
problem is that cooling is more efficient at high metallicities, which
translates into lower nebular temperatures and hence weaker [\ion{O}{2}]
and [\ion{O}{3}] visible-wavelength lines, with the primary cooling load
shifting to the far-infrared fine structure emission lines that cannot
be readily observed.  The result is that at high abundances the visible
wavelength spectra become increasingly insensitive to changes in \logoh.
In fact, some diagnostics, like the traditional R$_{23}$ line index,
effectively saturate at high metallicity \citep{kewley08}, making it
rather difficult to accurately measure even relative abundances.
Because we would like to verify the high estimated oxygen abundances and
we cannot reproduce the \citeauthor{tremonti04} abundance calculations,
which are based on a Bayesian statistical method, we measured \logoh\ in
our galaxies using two recommended methods from \citet{kewley08}.

Another possible source of systematic inaccuracy in the abundance
estimates is the treatment of extinction corrections.  The corrections
traditionally applied use the \ion{H}{1} Balmer decrement and assume a
simple uniform foreground obscuring screen model like that used for
stars to estimate $A_V$ and correct the other emission lines.  It is
expected, however, that extinction towards extended sources like
\ion{H}{2} regions is better described as a clumpy screen, for example
the turbulent screen models of \citet{fischera05}.  In these, use of a
simple screen tends to systematically {\it overestimate} the extinction
correction for the \ion{O}{2}\,$\lambda\lambda$3727,29\AA\ emission
line, leading to a systematic {\it underestimate} of the gas-phase
oxygen abundance.  In all of our galaxies the \ion{H}{1} Balmer
decrement measurements are consistent with low $A_V$ ($\lesssim 0.5$) for a
simple screen extinction model, so this is not a big effect compared to
other sources of measurement error given the attenuation curves in
\citet{fischera05}.

\citeauthor{kewley08} have measured the gas-phase metallicities in star
forming galaxies from SDSS using ten different methods, including that
of \citet{tremonti04}; they also provide average relations relating each
pair of methods.  While no one metallicity estimate is strictly
believable---i.e., the {\em true} Oxygen-to-Hydrogen ratio---the {\em
relative} measurements are generally robust \citep{kewley08}.  Using the
Data Release 6 SDSS spectra \citep{adelman08}, we subtracted the
underlying stellar continuum using the STARLIGHT program
\citep{fernandes05}.  We then calculated \tlogoh\ using the revised
\citet{kewley02} method given in Equation~(A3) of \citet{kewley08},
\begin{eqnarray}\label{eqn:kd02}
\nonumber \log(\mbox{[\ion{N}{2}]}/\mbox{[\ion{O}{2}]}) & = & 1106.8660
- 532.1451Z + 96.37326Z^2 \\
& & - 7.8106123Z^3 + 0.32928247Z^4,
\end{eqnarray}
where $Z \equiv 12 + \log(\mbox{O}/\mbox{H})$.  Like
\citeauthor{kewley08}, we found the roots of this equation using the
\texttt{fz\_roots} program in IDL.  We also measured the metallicity
using the ``O3N2'' method of \citet{pettini04}, as recommended by
\citeauthor{kewley08}, where 
\begin{equation}\label{eqn:pp04}
12 + \log(\mbox{O}/\mbox{H}) = 8.73 - 0.32\times 
  \log\left(\frac{[\mbox{\ion{O}{3}}]\lambda 5007/\mbox{H}\beta}{\mbox{[\ion{N}{2}]}\,\lambda 6584/\mbox{H}\alpha} \right).
\end{equation}
We choose to not use the R$_{23}$ diagnostic for measuring \tlogoh\
because it is known to be difficult to calibrate at these abundances
\citep{kewley02}.  In particular, while the seventeen $z > 0.024$
galaxies in our main sample span $\sim 0.2$~dex on the
\citeauthor{tremonti04} scale, they span $\sim 0.6$~dex when their
metallicities are calculated using the R$_{23}$ methods of either
\citet{mcgaugh91} or \citet{zaritsky94}, reflecting the fact that the
R$_{23}$ parameter essentially saturates at high oxygen abundances
\citep{bresolin07}.

\begin{figure}
\plotone{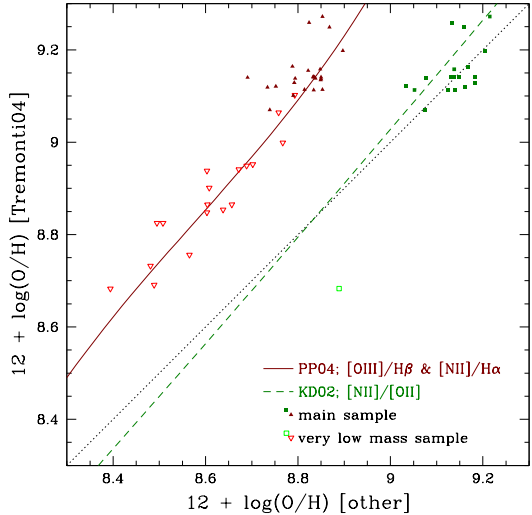}
\caption{\label{fig:convZ} Comparisons of oxygen abundances of
  low-luminosity mass--metallicity outliers measured with different
  methods as compared to mean relations.  The red triangles denote
  abundances measured with the method of \citet{pettini04}; the green
  squares are metallicities from revised \citet{kewley02} method from
  \citet{kewley08}.  The red solid line and green dashed lines are the
  conversion from the \citet{pettini04} method and the \citet{kewley02}
  method to the \citet{tremonti04} metallicity respectively; the black
  dotted line indicates equal measurements.  Conversions are from
  \citet{kewley08}; see \S\,\ref{sec:metal} for a more in depth
  discussion.}
\end{figure}

The oxygen abundances for our 41 galaxies plotted against these adopted
metallicities in Figure~\ref{fig:convZ}.  The dotted line indicates
equal measurements; the other two lines are the empirically determined
relations from \citeauthor{kewley08}.  The galaxies in our sample fall
preferentially above the relation for the \citeauthor{pettini04} method,
but surprisingly those galaxies for which it is measurable fall {\em
below} the relation for the \citeauthor{kewley02} method.  One possible
explanation for this discrepancy is that because [\ion{O}{2}] and
[\ion{N}{2}] are widely separated in wavelength, a misestimation of the
extinction or a poorly fitted continuum (especially for the weak
[\ion{O}{2}] emission line) can lead to underestimated abundances.
Specifically, there is often a degeneracy when fitting the stellar
continuum between the extinction and the stellar population.  While the
continuum fitting for our spectra are fairly good (i.e., they exhibit
low residuals after the stellar template is subtracted), we tested the
sensitivity to the continuum level at [\ion{O}{2}]\,$\lambda 3727$\AA\
by calculating the \citeauthor{kewley02}\ metallicity when the continuum
is over- and underestimated; the mean of these two extreme metallicities
is still systematically larger than expected by the mean relation,
implying that the shift is not due to continuum fitting error.  We
quantify the differences between the \citeauthor{pettini04} and
\citeauthor{kewley02} methods by examining the product of the vertical
displacement of each galaxy (for an optimally fit continuum) from the
two \citeauthor{kewley08}\ relations; only one galaxy in the main sample
has a significantly high \citeauthor{tremonti04} metallicity relative to
both of the other indicators.  However, this \tlogoh\ was still high
enough that even when the \citeauthor{tremonti04}\ metallicity was
replaced with the one expected from either relation, the galaxy still
passes the cuts in both the $M_g$- and M$_{\star}$-metallicity planes.
While the $z<0.024$ galaxies do not have measurable [\ion{O}{2}] (and
hence we cannot use the \citeauthor{kewley02}\ diagnostic for them), the
low redshift sample occupies a similar part of the
\citeauthor{tremonti04}\ versus \citeauthor{pettini04}\ parameter space
as the galaxies in the main sample.  For the very low-mass sample of
galaxies, out of the 22 galaxies which passed our visual inspection
test, we excluded 6 because they were more than $0.1$~dex above the
\citeauthor{pettini04}\ relation from \citeauthor{kewley08}; this is
actually a more stringent cut than used for the main sample because
several of the galaxies in the main sample with $>0.1$~dex deviations
from the \citeauthor{pettini04}\ relation have compensating negative
deviations relative to the \citeauthor{kewley02}\ scale. We interpret
these results to mean that the 41 galaxies we have identified are true
high-metallicity low-mass outliers from the mass--metallicity relation.

\begin{figure}
\plotone{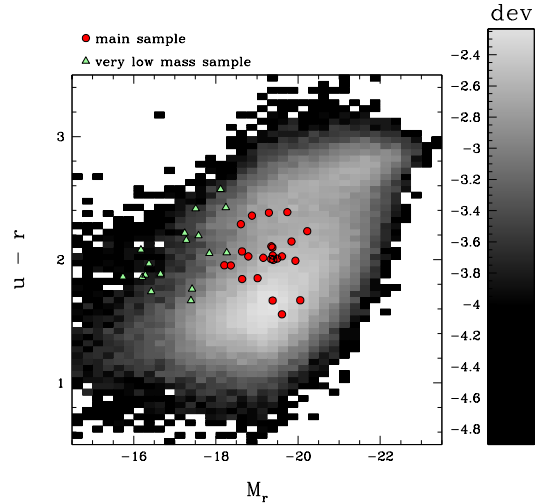}
\caption{\label{fig:cmd} High-metallicity low-mass galaxies on the
  color--magnitude diagram ($u-r$ vs.\ $M_r$) from the flux-limited $z <
  0.06$ sample of \citet{unterborn08}.  The color scale on the right
  shows the $\log f$ of each bin, where $f$ is the fraction of galaxies
  in that bin; bin sizes are $0.05$~mag in $u-r$ and $0.25$ in $M_r$.
  The red circles denote the 24 galaxies from the main sample and the
  green triangles denote the 17 galaxies in the very low mass sample. }
\end{figure}

\section{Discussion}\label{sec:disc}
The 41 dwarf galaxies selected as described in \S\,\ref{sec:outliers}
are surprisingly non-pathological.   They have undisturbed stellar
morphologies (see Figure~\ref{fig:images}).  Furthermore, given that by
selection these galaxies have both low luminosities and high
metallicities and that there is a trend for redder, brighter galaxies to
have higher oxygen abundances (see, e.g., \citealt{cooper08}) these
metal-rich dwarfs do not occupy unexpected region of the
color--magnitude diagram (see Figure~\ref{fig:cmd} and also
\S\,\ref{sec:trans}).  How, then, did they come to have such high oxygen
abundances?  One obvious possibility is that these metallicities are due
to an environmental effect, but as we describe in \S\,\ref{sec:enviro},
the galaxies in our sample are non-interacting and rather isolated.
Though several models predict that these galaxies should have high
specific star formation rates, we show in \S\,\ref{sec:sfr}, that these
galaxies have normal star formation rates for their masses.  We explain
in \S\,\ref{sec:gasfrac} why we predict that these metal-rich dwarf
galaxies should have relatively low gas fractions, and we discuss the
implications of this prediction in the broader context of
transition-type dwarf galaxies in \S\,\ref{sec:trans}.

\subsection{Environment}\label{sec:enviro}
While the origin of scatter in the mass--metallicity relation is
unknown, one popular proposal is that environment affects metallicity.
Recently, \citet{michel08} studied close pairs of star-forming galaxies
in SDSS with projected separations of $< 100\,$kpc and radial velocity
separations of $< 350\,$km\,s$^{-1}$.  Their results indicate that for
minor mergers, the less massive galaxy is likely to be preferentially
more metal rich than predicted by the mass--metallicity relation.
Likewise, \citet{cooper08}, after accounting for correlations within the
color--magnitude diagram, find a strong positive correlation between
metallicity and overdensity for $0.05 < z < 0.15$ SDSS-selected
star-forming galaxies that can account for up to $\sim 15$\% of the
scatter in the mass--metallicity relation.  In light of these results,
we searched for nearby neighbors of the seventeen $z > 0.024$ main
sample metal-rich dwarf galaxies in a cylindrical volume of depth $\pm
1000$\,km\,s$^{-1}$ and projected radius of 1~Mpc.  We find that our
galaxies are relatively isolated; none of them would have made it into
the \citet{michel08} sample.  Seven out of these seventeen galaxies in
our sample have no neighbors within this volume.  Of the 10 remaining
galaxies, four have no neighbors within 500\,km\,s$^{-1}$ and 500\,kpc,
and only two have any neighbors within 500\,km\,s$^{-1}$ and 100\,kpc.
One of these two seems to be on the outskirts of a nearby cluster, but
it is unclear whether or not it is physically associated with the
cluster (and there is only one small, faint, non-star-forming galaxy
within the \citeauthor{michel08}\ volume).  The other galaxy with a
nearby neighbor is $\sim 400$\,km\,s$^{-1}$ and 85\,kpc from a
less-massive relatively metal-poor ($12 + \log[\mbox{O}/\mbox{H}]
\approx 8.6$) star-forming galaxy.  We therefore conclude that
interactions with neighbors do not explain the observed high abundances
of our main sample of $\log\mbox{M}_{\star}\sim 9.5$ galaxies.

The galaxies in the very low mass sample (as well the $z < 0.024$
galaxies in the main sample), however, are generally at low enough
redshift that an automated search for neighbors in velocity and
projected distance space is difficult.  Like the galaxies in the main
sample, none of the lower-mass galaxies are in obviously interacting
systems.  Only one has an clear companion; it is a $\log\mbox{M}_{\star}
= 9.13$ galaxy which appears to be a satellite of a non-starforming
companion 33\,kpc and $\sim 120$\,km\,s$^{-1}$ away.  Also, while two of
the seventeen galaxies in the very low mass sample may be in the Virgo
cluster, clearly a rich environment cannot explain the high oxygen
abundances observed in all of these galaxies.

\subsection{Star Formation Rates}\label{sec:sfr}
\begin{figure}
\plotone{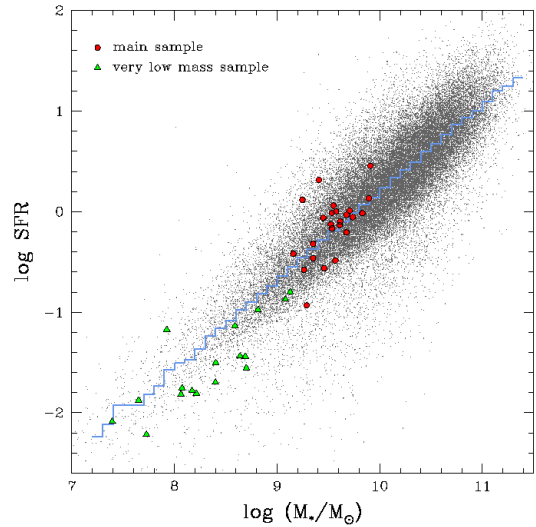}
\caption{\label{fig:sfrm} Star formation rate vs.\ stellar mass.  The
  grey points are galaxies from the \citet{tremonti04} sample with SDSS
  magnitude errors $<0.1$~mag, the galaxies in our main sample are
  denoted with red circles, and the galaxies in our very low mass sample
  are marked with green triangles.  The blue histogram denotes the median
  star formation rate in bins of $\log\mbox{M}_{\star}$\ of width
  0.1\,dex. The star formation rates estimates are from
  \citep{brinchmann04} and the stellar mass estimates are from
  \citet{tremonti04} and \citet{kauffmann03}. }
\end{figure}

\citet{dalcanton07} finds that for a galaxy to have a low metallicity,
it must have both a low star formation rate and a high gas fraction; she
argues that the observed low star formation efficiency in galaxies with
circular velocities $\lesssim 120$\,km\,s$^{-1}$ can explain the
mass--metallicity relation.  \citet{koppen07} similarly suggest that the
mass--metallicity relation could be due to a mass--star~formation rate
relation: a galaxy with a lower rate of star formation will have
relatively fewer massive stars, and therefore a lower oxygen abundance.
In either of these scenarios, we might expect the mass--metallicity
outliers to have high star formation rates for their masses.  On the
other hand, \citet{ellison08a} find that at low stellar masses, galaxies
with higher specific star formation rates tend to have {\em lower}
oxygen abundances.  As Figure~\ref{fig:sfrm} shows, the galaxies in the
main sample do not have preferentially high or low instantaneous
specific star formation rates, while typical star formation rates for
the the very low mass sample galaxies are $\sim 0.3$\,dex {\em lower}
than expected given their masses.

\subsection{Effective Yields and Gas Fractions}\label{sec:gasfrac}
So why do these galaxies have such high oxygen abundances?  Consider a
closed box star forming system (i.e., a system with no gas inflow or
outflow).  The metallicity $Z\equiv\;$(mass of metals in gas
phase)/(total gas mass) is
\begin{equation}\label{eqn:closedbox}
Z = y\ln\left(\frac{1}{f_{\mbox{\scriptsize gas}}}\right),
\end{equation}
where $f_{\mbox{\scriptsize gas}}\equiv M_{\mbox{\scriptsize
gas}}/(M_{\star}+M_{\mbox{\scriptsize gas}})$ is the gas fraction and
$y\equiv\;$(mass of metals in gas)/(mass of metals in stellar remnants
and main sequence stars) is the metal yield.  An immediately striking
aspect of Equation~\ref{eqn:closedbox} is that there is no implicit
dependence on the total galaxy mass; the deviations from a universal
metallicity\footnote{While the $Z$ in Equation~\ref{eqn:closedbox} is
{\em not} the same as \tlogoh---it is a mass ratio of {\em all} metals
rather than the abundance ratio of one element relative to
Hydrogen---the same arguments still qualitatively hold for observed
abundances.} observed via the mass--metallicity relation are presumably
then due to the fact that galaxies are not scaled closed box versions of
one another: variations of the yield with mass, a dependence on the gas
fraction with mass, or a combination of these effects plays a role.  For
a galaxy to have a higher metallicity than other galaxies of the same
mass, Equation~\ref{eqn:closedbox} tells us that it must have either a
relatively high yield or a relatively low gas fraction.
Observationally, the yield---which depends on both star formation
physics as well as gas inflow and outflow---is a difficult quantity to
measure; one popular way to address this problem is to define an
effective yield,
\begin{equation}\label{eqn:yeff}
y_{\mbox{\scriptsize eff}} \equiv \left[\frac{Z_{\mbox{\scriptsize
	gas}}}{\ln (1/f{\mbox{\scriptsize gas}})}\right],
\end{equation}
as the yield the galaxy {\em would} have were it actually a closed
system.  Given the galaxy's metallicity and gas fraction, one can then
easily calculate $y_{\mbox{\scriptsize eff}}$.  A common explanation for
the mass--metallicity relation is that the effective yield is positively
correlated with galaxy mass, so that lower mass galaxies are able to
preferentially lose metals to the intergalactic medium via winds because
of their relatively shallow potential wells \citep[see e.g.,][and
references therein]{larson74, finlator08}.  In particular, 
\citet{tremonti04} tested this idea by comparing effective yields and
baryonic masses.  However, because when estimating the effective yield
from Equation~\ref{eqn:yeff}, \citeauthor{tremonti04}\ use the measured
\tlogoh\ values, outliers in the mass--metallicity relation are
practically guaranteed to be outliers in the effective yield--baryonic
mass relation.

Regardless, it is obvious from Equation~\ref{eqn:closedbox} that our
metal-rich dwarfs must either have unusually high yields or unusually
low gas fractions for their masses.  Galaxy yields can be affected by
three processes: star formation, metal-deficient gas inflows, and
metal-rich gas outflows.  As shown in Figure~\ref{fig:sfrm} and
discussed in \S\,\ref{sec:sfr}, we find that the star formation rates
for this population of galaxies are consistent with those of other
galaxies of similar masses.  \citet{dalcanton07} has shown that
metal-poor gas inflow is insufficient to explain the typically low metal
abundances of low-mass galaxies; we therefore conclude that a lower
inflow rate is insufficient to explain the higher abundances of our
galaxies.  It is possible that these galaxies are less effective at
driving outflows than other dwarfs.  For example, \citet{ellison08a}
find that, at fixed mass, galaxies with smaller half-light radii tend to
have higher abundances; this picture is consistent with the idea that it
is more difficult to drive winds from deeper potential wells.
(\citealt{tremonti04}, on the other hand, find no correlation with how
concentrated a galaxy's light is and its oxygen abundance.)  While the
galaxies in our main sample do tend to have small radii for their masses
(half-light radii of less than 2\,kpc), the typical \tlogoh\ for
galaxies of similar masses and radii is still $\sim 0.3$\,dex lower
(about $2\sigma$) than that of our galaxies.  Furthermore, the very low
mass sample galaxies do not have preferentially small radii for their
masses, leading us to conclude that while small radii may be a
contributing factor to why some of these galaxies have been able to
retain their metals, size alone does not tell the whole story. 

\citet{dalcanton07} calculated that enriched gas outflows can only
severely decrease a galaxy's effective yield if the gas fraction is
sufficiently high; that is, for low gas fractions, even a very strong
outflow cannot drastically decrease the effective yield---and thus
measured abundance.  Likewise, if a galaxy has a relatively low gas
fraction, then only a small amount of pollution is needed to enrich the
gas and cause the measured abundance to be high.  We therefore predict
that these mass--metallicity outliers have anomalously low gas masses
relative to other isolated galaxies of similar luminosities and
star-formation rates.  There are unfortunately no \ion{H}{1} data in the
literature for our galaxies that we can call upon to lend observational
support to this prediction.  However, \citet{lee06} have found that, in
a sample of 27 nearby dwarf irregular galaxies, the gas-phase oxygen
abundance is negatively correlated with the \ion{H}{1}-measured
gas-to-stellar mass ratio, which lends observational credence to our
expectation.

\subsection{Transitional Dwarf Galaxies}\label{sec:trans}
If these high-metallicity dwarf galaxies really do have relatively
little gas, then they should be rapidly approaching the end of their
star formation.  Specifically, these galaxies are likely to be
transitioning from gas-rich dwarf irregulars (dIrr) to gas-deficient
dwarf spheroidals (dSph) or the more massive dwarf ellipticals (dE).  In
general, so-called dIrr/dSph transitional dwarfs have similar
star-formation histories as their currently non-starforming dSph
cousins: both galaxy types typically have a mix of old and
intermediate-age stellar populations \citep{grebel03}.  In a study of
five nearby star-forming transitional dwarfs, \citet{dellenbusch07}
found that these relatively isolated galaxies seem to have unusually
high oxygen abundances for their luminosities, much like our sample of
mass--metallicity outliers.\footnote{All five of the
\citeauthor{dellenbusch07}\ galaxies are in the \citeauthor{tremonti04}\
sample; four were either too high mass or too low metallicity (as
measured by \citeauthor{tremonti04}) to be significant outliers by our
definitions. The final galaxy (IC~745) was excluded from our sample
because its measured \citet{pettini04} abundance when converted to the
\citeauthor{tremonti04}\ scale is 0.22~dex above the \citet{kewley08}
relation (as discussed in \S\,\ref{sec:metal}), highlighting the point
that while our sample is rather pure, it is almost certainly
incomplete.}  \citet{grebel03} stress that the only difference between
dSph galaxies and the transitional dIrr/dSph galaxies is the absence of
star formation and of gas in dSph galaxies.  While much of this
conclusion is based on galaxies in higher-density environments than ours
are (i.e., the \citet{grebel03} dwarf galaxies are mostly in the Local
Group), we find it likely that our galaxies are part of a similar
transition population: morphologically, they have smooth, undisturbed
profiles (see Figure~\ref{fig:images}), indicating a stronger
relationship to dwarf spheroidals than to the dwarf irregulars.  In
particular, for all of our galaxies we note a lack of the irregular
flocculent structures or spiral features often associated with dIrr
galaxies.

\begin{figure*}
\plotone{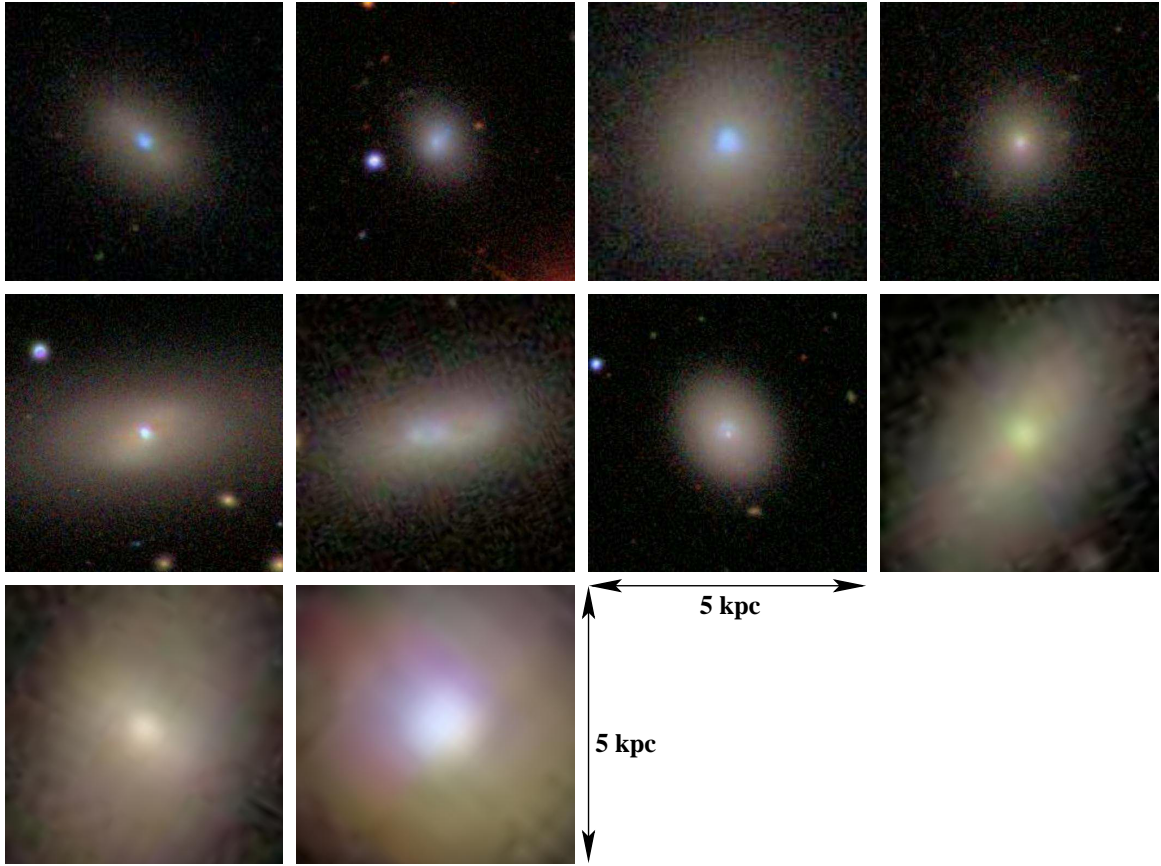}
\caption{\label{fig:blue} SDSS images of high-metallicity low-mass
galaxies with bright and/or blue cores, scaled to
$5\times5$\,kpc, with $M_B$ decreasing to the right and
down so that the faintest galaxy is in the upper-left corner. }
\end{figure*}

In fact, the only remarkable morphology any of the galaxies in our
sample displays are the bright---and often very blue---cores noticed in
ten of the galaxies.  These galaxies are shown in Figure~\ref{fig:blue}
and noted in Table~\ref{tbl:sample}.  Some of these cores appear to have
merely much higher surface brightnesses than their surroundings, but
some are decidedly blue: \citet{gu06} measure a difference of
$\Delta(g-r) \gtrsim 0.3$~mag~arcsec$^{-2}$ from the center of
IC~225\footnote{As mentioned in \S\,\ref{sec:lowm}, IC~255 did not pass
all of the cuts for our very low mass sample, and so we reinserted it
into the sample by hand.} relative to isophotes at $>5$\arcsec. These
bright cores are almost certainly not a signature of active galactic
nuclei; as mentioned in \S\,\ref{sec:metal}, these galaxies fall
squarely within the star-forming locus of the normal BPT line-ratio
diagnostic diagrams \citep{baldwin81,kewley06b}.  \citet{boselli08} find
that for similar transitional dwarf galaxies with blue centers in the
Virgo cluster (including VCC~1855, the least massive galaxy in our
sample with $\log [\mbox{M}_{\star}/\mbox{M}_{\odot}]\sim 7.4$), star
formation is limited to these central blue nuclei.
(\citeauthor{boselli08}\ attribute this centralized star formation to
the effects of ram pressure stripping preferentially removing gas from
the outer regions of the galaxies.  However, not all of the galaxies in
our sample with bright nuclei are in the kinds of rich environments like
the Virgo cluster where ram pressure stripping expected to play an
important role.)  Intriguingly (yet unsurprisingly), most of our
galaxies which appear to be approaching or on the red sequence (see
Figure~\ref{fig:cmd}) also have bright and/or blue cores, implying that
their colors are already dominated by their non-starforming regions.
Furthermore, having star formation limited to a relatively small region
of the galaxy supports the idea that there is relatively little fuel
available for forming stars, and thus that these galaxies are on their
way to becoming standard dwarf spheroidal or dwarf elliptical galaxies.

\section{Conclusions}\label{sec:conc}
We have identified a sample of 41 low-luminosity high--oxygen abundance
outliers from the mass--metallicity relation.
\begin{enumerate}
\item These galaxies are fairly isolated and none show any signs of
  morphological disturbance.  It is likely, however, that the handful of
  galaxies in this sample which are possible group or cluster members
  have disturbed {\em gas} morphologies, due to e.g., ram-pressure
  stripping.  However, environmental effects cannot account for the
  large gas-phase oxygen abundances observed in all of these galaxies.
\item Based on the effective yield considerations of \citet{dalcanton07}
  and various related observational results \citep{grebel03,
  dellenbusch07, lee06, boselli08}, we predict that these galaxies
  should have low gas fractions relative to other dwarf galaxies of similar
  stellar masses and star formation rates in similar environments but
  with more typical (i.e., lower) oxygen abundances.  Furthermore, in
  this scenario, the {\em stellar} metallicities should be consistent
  with that of other dwarfs of similar luminosity, i.e., the stellar
  metallicities should be lower than the measured gas-phase abundances.
\item These mass--metallicity outliers appear to be ``transitional''
  dwarf galaxies: because they are running out of star-forming fuel,
  they are nearing the end of their star formation and becoming typical
  isolated dwarf spheroidals.  This conclusion is most sound for the
  $\mbox{M}_{\star} < 9.1 M_{\sun}$ galaxies which are observed to have low
  specific star formation rates.  More data---specifically, gas
  fractions and stellar metallicities---are needed to help elucidate the
  situation for the higher mass galaxies.
\end{enumerate}

\acknowledgements 
We would like to thank Roberto Assef for adapting his
low-resolution template code for calculating $M_B$ for the SDSS sample
for us, Cayman Unterborn for letting us use his data and code for
generating the color--magnitude diagram in Figure~\ref{fig:cmd}, and
Sara Ellison for giving us access to some of the data from
\citet{ellison08a}.  We are grateful to the anonymous referee, David
Weinberg, Christy Tremonti, and Thorsten Lisker for helpful comments on
the text.  We would also like to thank Jos\'e Luis Prieto, Jeff Newman,
Mike Cooper, Kristian Finlator, John Moustakas, Henry Lee, Paul Martini,
and Todd Thompson for useful discussions and suggestions.

    Funding for the SDSS and SDSS-II has been provided by the Alfred
    P.\ Sloan Foundation, the Participating Institutions, the National
    Science Foundation, the U.S. Department of Energy, the National
    Aeronautics and Space Administration, the Japanese Monbukagakusho,
    the Max Planck Society, and the Higher Education Funding Council for
    England. The SDSS Web Site is \texttt{http://www.sdss.org/}.

    The SDSS is managed by the Astrophysical Research Consortium for the
    Participating Institutions. The Participating Institutions are the
    American Museum of Natural History, Astrophysical Institute Potsdam,
    University of Basel, University of Cambridge, Case Western Reserve
    University, University of Chicago, Drexel University, Fermilab, the
    Institute for Advanced Study, the Japan Participation Group, Johns
    Hopkins University, the Joint Institute for Nuclear Astrophysics,
    the Kavli Institute for Particle Astrophysics and Cosmology, the
    Korean Scientist Group, the Chinese Academy of Sciences (LAMOST),
    Los Alamos National Laboratory, the Max-Planck-Institute for
    Astronomy (MPIA), the Max-Planck-Institute for Astrophysics (MPA),
    New Mexico State University, Ohio State University, University of
    Pittsburgh, University of Portsmouth, Princeton University, the
    United States Naval Observatory, and the University of Washington.

The STARLIGHT project is supported by the Brazilian agencies CNPq, CAPES
and FAPESP and by the France-Brazil CAPES/Cofecub program.

\end{document}